# On the thickness of the double layer in ionic liquids

Anton Ruzanov[a], Meeri Lembinen[b], Pelle Jakovits[c], Satish N. Srirama[c], Iuliia V. Voroshylova[d,e],
M. Natália D.S. Cordeiro[d], Carlos M. Pereira[e], Jan Rossmeisl[f], and Vladislav B. Ivaništšev[a]*

In this study, we examined the thickness of the electrical double layer (EDL) in ionic liquids using density functional theory (DFT) calculations and molecular dynamics (MD) simulations. We focused on the $BF_4^-$ anion adsorption from 1-ethyl-3-methylimidazolium tetrafluoroborate ($EMImBF_4$) ionic liquid on the Au(111) surface. At both DFT and MD levels, we evaluated the capacitance–potential dependence for the Helmholtz model of the interface. Using MD simulations, we also explored a more realistic, multilayer EDL model accounting for the ion layering. Concurrent analysis of the DFT and MD results provides a ground for thinking whether the electrical double layer in ionic liquids is *one*- or *multi*-ionic-layer thick.

## Introduction

Since their rediscovery in the 1990s, the so-called room temperature ionic liquids (RTILs) have been studied as solvents with a unique combination of diverse physicochemical properties,[1–3] that make them captivating from fundamental and application points of view.[4–9] Within the following decades, electrode | RTIL interfaces have attracted considerable attention. Currently, the electrical double layer (EDL) at the electrode | RTIL interfaces is in focus of research on enhancing performance of energy storage and transformation in supercapacitors,[10,11] actuators,[8,12] batteries,[2,13] and solar cells.[14]

Significant progress in understanding of the interfacial processes occurring in the EDL has been made recently at the theoretical level,[15–20] by computational modelling,[21–27] and in experimental measurements.[28–43] Nevertheless, some authors argue whether the EDL in RTILs is *one*- or *multi*-ionic-layer thick. On the one hand, by vibrational Stark shifts and capacitance measurements, Baldelli concluded that the EDL in RTILs is effectively *one*-ionic-layer thick due to a single layer of counter-ions.[32,33] On the other hand, other authors considered a multilayer structure for interpretation of electrochemical impedance data.[34,36] In theory, an account for the innermost layer of counter-ions is crucial in modified Poisson–Boltzmann, mean spherical approximation, and Landau–Ginzburg-type continuum models.[17–19,44,45] Molecular dynamics (MD) simulations,[26,46–48] atomic force microscopy,[49] and X-ray spectroscopy[50–52] studies have ascertained that the EDL in RTILs indeed consists of alternating layers of anions and cations.

* Corresponding author. E-mail: vladislav.ivanistsev@ut.ee
[a] Institute of Chemistry, University of Tartu, Ravila 14a, 50411 Tartu, Estonia
[b] Institute of Physics, University of Tartu, Ostwald str. 1, 50411 Tartu, Estonia
[c] Mobile & Cloud Computing Laboratory, Institute of Computer Science, University of Tartu, J. Liivi 2, 50409 Tartu, Estonia
[d] Departamento de Química e Bioquímica, LAQV@REQUIMTE, Faculdade de Ciências, Universidade do Porto, Rua do Campo Alegre, 4169-007 Porto, Portugal
[e] Departamento de Química e Bioquímica, CIQ(UP), Faculdade de Ciências, Universidade do Porto, Rua do Campo Alegre, 4169-007 Porto, Portugal
[f] Department of Chemistry, University of Copenhagen, Universitetsparken 5, København 2100, Denmark

According to these studies, in the innermost layer, the counter-ions are in direct contact with the surface, templating the subsequent layers. Upon closer examination of MD simulations results, it appears that the EDL structure changes from *multi*- to *mono*layer upon variation of the surface charge.[25,27] Overall, it may be assumed that the *innermost layer* largely determines the interfacial properties, yet the extent remains unclear.[53]

How thick is the EDL in RTILs and does the innermost layer dominate in the overall potential-dependent multilayer EDL? We endeavoured to investigate the subject as the answer to these questions is essential for development of energy storage devices, especially supercapacitors and batteries.[1,2]

We focused on the adsorption[†] of $BF_4^-$ anions from 1-ethyl-3-methylimidazolium tetrafluoroborate ($EMImBF_4$) ionic liquid on the Au(111) surface using density functional theory (DFT) calculations and MD simulations. First, we examined the differences in the DFT and MD representation of the Helmholtz model of the Au(111) | $BF_4^-$ interface. Next, in comparison to the Helmholtz model, we explored a more realistic, multilayer Au(111) | $EMImBF_4$ interface accounting for the ion layering.

To the best of our knowledge, only Valencia *et al.*,[54–56] Klaver *et al.*,[57] and Plöger *et al.*[58] conducted similar DFT calculations to study the adsorption of RTILs on *uncharged* lithium, gold, aluminium, and copper surfaces. Differently, in this study, we investigated the adsorption of $BF_4^-$ on *charged* Au(111) surface.

The interfaces between imidazolium tetrafluoroborate RTILs and single crystal Au(111), Cd(0001), Bi(111) as well as polycrystalline gold and platinum surfaces were previously studied using cyclic voltammetry and electrical impedance spectroscopy techniques.[36–40,43,59–63] The measured capacitance dependence on potential is widely agreed to be determined by the adsorption of anions/cations, implying accumulation of ionic counter charge near the charged metal surface. For some ions, the formation of the ordered adlayers at single-crystal gold faces was observed by *in situ* scanning tunnelling microscopy.[43,64–67] Based on these findings, we assumed that an ordered layer of $BF_4^-$ describes the Au(111) | $EMImBF_4$ interface at anodic potentials.

## Interface models and computational methods

### Interface models

As a rough approximation of a positively charged electrode immersed into an EMImBF$_4$ ionic liquid, we constructed a set of Au(111) | BF$_4^-$ interface configurations representing the Helmholtz model. In this model, the surface charge is compensated by a layer of counter-ions at an average distance $d$ from the metal surface. This model represents a simple parallel plate capacitor.

In the DFT calculations, the coverage ($\theta$) ranged from 1/20 to 1/2 of BF$_4^-$ anions per surface gold atom in the unit cell of variable size. Three layers of gold atoms in total formed the slab representing the Au(111) surface. Fig. 1a shows Au(111) | BF$_4^-$ interface model at $\theta$ = 1/3. Only the first upper Au layer was allowed to relax, while the two bottom layers were kept fixed in their bulk positions. As a starting guess for the electronic structure, a varying number of BF$_4^\bullet$ radicals were placed on the neutral Au(111) surface. The charge of the radicals spontaneously decreased during relaxation, turning the BF$_4^\bullet$ radicals into BF$_4^-$ ions. Consequently, the interface became polarised, and the electric field set up between the charged gold surface and adsorbed ions.

Within the Helmholtz model framework, we looked only at the adsorption of anions on the Au(111) surface, in the absence of cations. This divide-and-conquer approach is a reasonable first step towards more complex models. According to Ref. [53], the current model is the simplest "1D" representation of the EDL in RTILs. Recent MD simulations results reveal that the EDL structure can indeed be reduced to "1D" at a certain surface charge when a monolayer of counter-ions at a charged surface is formed.[27,68] Thus, we thoughtfully utilised the Helmholtz model not only to test its limits but also to verify the concept of the monolayer formation.

In the MD simulations of the Helmholtz model, a variable number of BF$_4^-$ anions (from 1 to 112) were put into contact with a fixed Au(111) slab. The cell size was 4.04×4.00 nm$^2$ and consisted of 224 gold atoms. Each anion had a total charge of $-e/\sqrt{2}$, and each gold atom had a fixed point charge required to compensate the overall ionic charge. The studied range of the surface charge density varied from 1 to 80 μC/m$^2$ corresponding to the coverage of 1/2.

In the more realistic MD simulations, the initial configurations were constructed using the PACKMOL package,[69] by inserting 288 cations and 288 anions of EMImBF$_4$ at random positions between two golden slabs to form the final simulation cell, with dimensions of 2.98 nm × 2.95 nm × 11.36 nm. The golden slab of Au(111) was setup using 480 gold atoms with the help of atomic simulation environment (ASE).[70] The golden slabs were fixed in positions during all simulations. The polarisation was realised by applying an electric field in the $z$-direction of the simulation cell. According to our preliminary tests, this approach is equivalent to the assigning of point charges (as in the case of the Helmholtz model), but it is computationally more efficient.

### Density functional theory calculations

All DFT calculations were performed with the ASE interface using the revised Perdew–Burke–Ernzerhof (RPBE) exchange-correlation functional and projector augmented wave (PAW) method as implemented in the real-space grid code GPAW.[70–72] A van der Waals (vdW) correction proposed by Tkachenko and Scheffler was applied on top of the RPBE functional.[73] Wave functions, potentials, and electron densities were represented on grids with a spacing of approximately 0.16 Å. Brillouin-zone integrations were performed using an $a$×$b$×1 Monkhorst–Pack $k$-point sampling grid, where $a$ and $b$ equalled 2 or 4 depending on the size of the surface lattice cell. Molecules were computed in a large non-periodic cell while the surface lattice cell was repeated periodically in the surface plane to create an infinite metal slab. The energy convergence on $k$-points and $h$-spacing was tested on the modelled slabs. Dipole correction was employed in the perpendicular direction to the slab to decouple two adjacent images electrostatically. The structural optimisations were performed with a convergence criterion of 0.05 eV/Å for atomic forces.

The starting geometry for EMIm$^+$–BF$_4^-$ ionic pair and lattice parameters for EMImBF$_4$ crystal were taken from supporting information in Refs. [74–76] and optimised with RPBE+vdW. The dissociation energy −344 kJ/mol for EMIm$^+$–BF$_4^-$ ionic pair agrees with the post-Hartree–Fock results.[74,77] The energy of EMImBF$_4$ crystal dissociation into single ions is 161 kJ/mol lower. It is in reasonable agreement with the experimentally determined value for EMImBF$_4$ liquid evaporation (135–149 kJ/mol at 298 K).[78,79] EMImBF$_4$ crystal structure optimisation was performed on Amazon EC2 public cloud,‡ using the Desktop to Cloud Migration (D2CM) tool.[80,81]

The binding energy of BF$_4^-$ in the modelled systems was expressed relative to the potential energy of BF$_4^-$ in vacuum, and corrected by the BF$_4^\bullet$ adiabatic electron affinity (EA):[82]

$$E_{\text{surf}}(\text{BF}_4^-) = [E(N,n) - E(N,0) - nE(\text{BF}_4^-) - n\text{EA}(\text{BF}_4^\bullet)]/n, \quad (1)$$

where $n$ and $N$ are the numbers of ions and surface metal atoms in the simulated cell, and $E(N,0)$ and $E(N,n)$ are the potential energies of the bare Au(111) surface and the charged Au(111) surface with $n$ BF$_4^-$ species in the cell. The adiabatic electron affinity of BF$_4^\bullet$ calculated with RPBE+vdW in this work (634 kJ/mol) agrees well with the value of 649 kJ/mol, calculated by Gutsev et al. at CCSD(T)[83] level of theory.

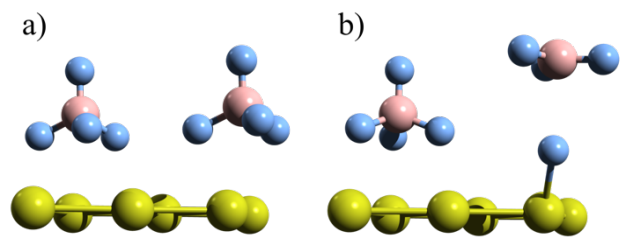

**Figure 1.** a) Au(111) | BF$_4^-$ interface model at surface coverage $\theta$ = 1/3. b) BF$_4^-$ reorients and spontaneously dissociates at the same coverage. Only the upper, relaxed layer of gold is shown.

The $E(BF_4^-)$ can be directly related to the Madelung energy of $BF_4^-$ in the EMImBF$_4$ crystal ($E_{cr}$) which may be considered as an approximation to the electrochemical potential of the anion in the RTIL. The formation energy of a vacancy in EMImBF$_4$ crystal is expressed as follows:[84]

$$E_{cr}(BF_4^-) = E(EMImBF_4) - E(BF_4^-) - E(EMIm^+), \quad (2)$$

where $E(EMImBF_4)$ is the potential energy of the EMImBF$_4$ crystal, and $E(EMIm^+)$ is the potential energy of EMIm$^+$ in vacuum. Once an anion leaves the crystal, it can be deionised to $BF_4^{\bullet}$ and then can adsorb on the Au(111) surface.

For the Au(111) | BF$_4^-$ interface, the integral free energy change per surface metal atom ($\Delta G_{int}$) was defined as[85–87]

$$\Delta G_{int} \approx [nE_{surf}(BF_4^-) - nE_{cr}(BF_4^-)]/N. \quad (3)$$

In this form, the integral free energy change serves as a measure of the BF$_4^-$ affinity towards the surface compared to the affinity towards the EMImBF$_4$ crystal. It was calculated using the values consistent with the forces presented in the models and neglecting the contribution of entropy terms non-presented in the models. This should be a sufficiently reliable estimate of the interfacial free energy, as on the transition from a real RTIL to an interface the ions remain in glass-like phase. Thus, the entropy term ($T\Delta S$) should be much smaller than the enthalpy change ($\Delta H$), determined by strong Coulomb interaction in RTILs.

The *integral* capacitance was determined in four ways from the integral free energy, work function, ionic charges, and interfacial dipole moment.

Firstly, using the classical relation:

$$C_G = 2\Delta G_{int}/\Delta U^2, \quad (4)$$

where $\Delta G_{int}$ is equal to the energy stored in an ideal capacitor which, in our case, is set up by BF$_4^-$ ions and the counter charge on the metal surface. Here, $\Delta U = U - U_{pzc}$ is the electrode potential calculated from the work function ($U = W_e/e$), and $U_{pzc}$ is the potential of zero charge (PZC). Here $e$ is the elementary electronic charge. The $U_{pzc}$ was set to be equal to calculated work function of the Au(111) surface (5.08 eV), which is slightly lower than the experimental value of 5.26 eV.[88]

Secondly, taking into account that each anion brings a charge of $q$ to the surface:

$$C_\theta = q \cdot e/A \cdot \theta/\Delta U, \quad (5)$$

where $A$ is the area of the unit cell, and the ionic charge ($q$) was obtained by the density derived electrostatic and chemical (DDEC) method.[89,90]

Thirdly, using the interfacial dipole moment ($\mu$):

$$C_\mu = q \cdot e \cdot \varepsilon_0/\mu, \quad (6)$$

where $\varepsilon_0$ is the permittivity of vacuum.

Finally, assuming that the system is a parallel plate capacitor

$$C_H = \varepsilon\varepsilon_0/d, \quad (7)$$

where $\varepsilon$ is the high-frequency dielectric constant of 2.0 (typical for RTILs [91]), and $d$ is the distance from the position of the nearest layer of Au nuclei to the layer of B nuclei. Eq. 7 is derived based on the Helmholtz model assumptions.

**Molecular dynamics simulations**

All MD simulations were carried out using the GROMACS 2016.1 simulation package[92] and NaRIBaS scripting framework[93] at temperatures of 300 K. The parameters of Lennard-Jones potential for gold reported by Heinz *et al.* were utilised.[94] OPLS-AA force field was used for EMImBF$_4$.[95]

For the initial relaxation of EMImBF$_4$, all systems were subjected to the steepest descent minimisation. Then, the 20 ns equilibration run, followed by the 2 ns of system polarisation, were performed. Finally, a production run of 10 ns was accomplished.

A coupling constant for a V-rescale thermostat, used throughout all calculations to maintain constant the temperature, was 0.5 ps.[96] Due to the slab-like geometry of studied systems, the periodic boundary conditions were applied only in $x$ and $y$ directions. The Verlet leapfrog algorithm was used to integrate the equations of motion, with a time step of 1 fs.[97] The short-range non-bonded Coulomb and Lennard-Jones interactions were computed with a 1.45 nm cut-off distance. The same cut-off distance was used for the short-range neighbour list. The corrections for the Coulomb interactions beyond the cut-off were performed using the particle mesh Ewald method with interpolation order of 6 and 0.1 nm spacing of the grid points in the reciprocal space.[98] The 3dc Ewald geometry was used, *i.e.* the force and potential corrections were applied in the $z$-dimension to produce a pseudo-2D summation. The short-range interaction lists were updated every 40 steps, using a grid-based method. The dielectric constant was chosen to be 2.0. Constraints were enforced on all bond lengths using the LINCS algorithm.[99] Trajectory data were written at every 5 ps and later analysed using GROMACS inbuilt tools and our codes, whenever the respective analysis tool was unavailable in GROMACS.

The *integral* capacitance was determined using the values of the calculated electrostatic potential drop ($U$) and the surface charge density:

$$C = \sigma/\Delta U \quad (8)$$

where $\sigma$ denotes the surface charge density and $\Delta U = U - U_{pzc}$ obtained by integrating the Poisson's equation (see Ref. [100] for details). The screening of the external field by the gold atoms was characterised by an effective position of the image plane. The position was fixed at 0.135 nm from the nearest layer of Au nuclei, which is slightly further than half an Au(111) interlayer spacing (0.118 nm). This position value was taken from Ref. [101], at higher surface charges.

**Table 1.** For different surface coverage ($\theta$) the values of the adsorption energy ($E_{ads}$ / kJ mol$^{-1}$) were evaluated from the results of calculations using RPBE and vdW-DF functionals, ionic charges ($q$ / $e$) were obtained using DDEC method,[89,90] whereas the distance ($d$ / Å) was calculated from the position of the nearest layer of Au nuclei to the layer of B nuclei.

| Adsorbate | $\theta$ | $-E_{ads}$ / kJ mol$^{-1}$ | $-q$ / $e$ | $d$ / Å |
|---|---|---|---|---|
| $BF_4^-$ | 1/3 | 237 | 0.41 | 2.94 |
| $BF_4^-$ | 1/4 | 268 | 0.41 | 2.99 |
| $BF_4^-$ | 1/6 | 281 | 0.49 | 3.01 |
| $BF_4^-$ | 1/12 | 301 | 0.62 | 3.11 |
| $BF_4^-$ | 1/20 | 308 | 0.67 | 3.13 |
| $F^\bullet(BF_3)$ | 1/3 | 248 | 0.36 | |
| $F^\bullet$ | 1/3 | 221 | 0.34 | |
| $F^\bullet$ | 1/4 | 213 | 0.34 | |
| $F^\bullet$ | 1/6 | 224 | 0.37 | |
| $F^\bullet$ | 1/12 | 225 | 0.38 | |
| $F^\bullet$ | 1/20 | 232 | 0.39 | |

## Results and discussion

### Model Au(111) | $BF_4^-$ interface

To obtain a qualitative comparison of the preferred orientation of a single $BF_4^-$ ion, the usual adsorption sites on the surface were considered: FCC and HCP hollow, bridge and top sites. At 1/20 coverage, the FCC and HCP hollow sites were found to be the most stable adsorption sites with a negligible energy difference. At the same time, the translational movement of $BF_4^-$ from an FCC to an HCP hollow site requires overcoming an energy barrier of 11 kJ/mol. Up to coverage of 1/3, the orientation of anion with three fluorine atoms pointing towards the surface is the most favourable. At higher coverages, the re-orientation of anions can happen during the geometry optimisation.

Fig. 2 demonstrates the dependence of the integral energy ($G_{int}$) of $BF_4^-$ anions on the relative electrode potential squared ($\Delta U^2$). A linear dependence is seen, which means that the modelled Au(111) | $BF_4^-$ interface behaves as a parallel plate capacitor. Yet, the results also indicate a strong potential dependence on the orientation of $BF_4^-$ ions. As it follows from the calculations, the formation of the √3×√3 ordered adlayer (Fig. 1a, $\theta$ = 1/3) occurs at 4.6 V relative to the PZC (Fig. 2, Table 2). However, the formation of the √3×√3 adlayer with half of the anions flipped and dissociated (Fig. 1b) can take place at a considerably lower potential of 3.5 V (Fig. 3).

As it can be seen in Fig. 3, at $\theta$ = 1/3, the physical adsorption energy difference between the undissociated and the dissociated structures is relatively small (11 kJ/mol), yet the potential difference is pronounced (1.1 V). On the one hand, the potential is directly related to the interfacial dipole moment, in a direction perpendicular to the surface, which is apparently determined by the orientation of species ($BF_4^-$, $F^-$, $BF_3$). On the other hand, the adsorption energy results from the lateral repulsion among the species, which is less sensitive to their orientation.

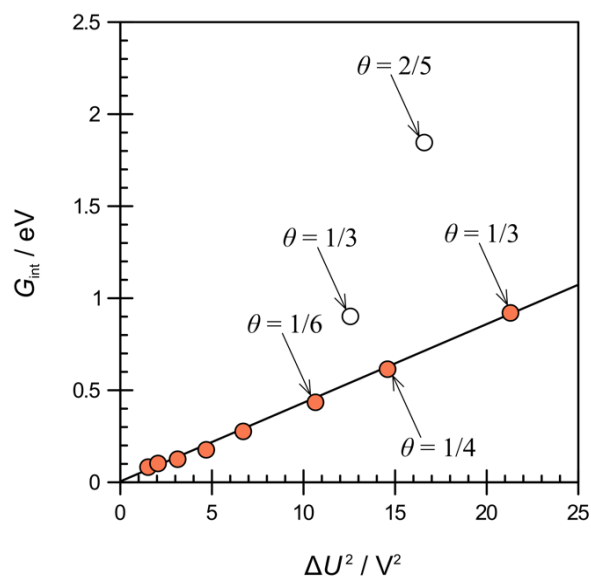

**Figure 2.** Dependence of the integral energy ($G_{int}$) of $BF_4^-$ anions (●) on the relative electrode potential squared ($\Delta U^2$). Blank markers (○) indicate $BF_4^-$ dissociation to $BF_3$ + $F^\bullet$. Surface coverage ($\theta$) is labelled with arrows. The slope corresponds to the differential capacitance value of 6 µF/cm$^2$.

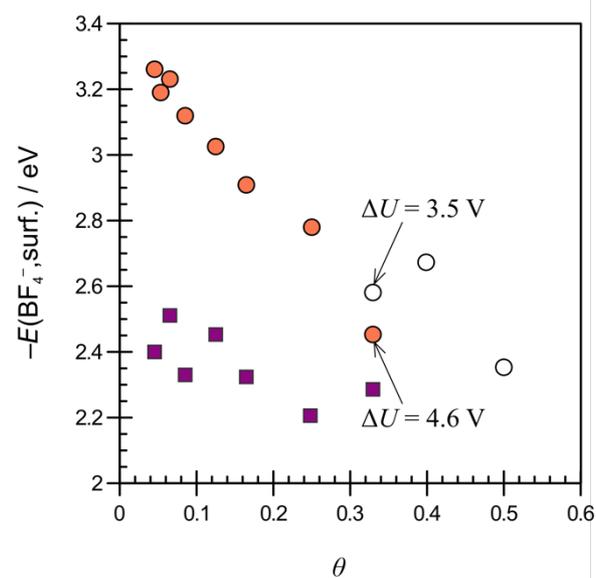

**Figure 3.** Dependence of the $BF_4^-$ (●) and $F^-$ (■) adsorption energies ($E_{ads}$) on the surface coverage. At a high coverage, $BF_4^-$ is oxidized and dissociates into $BF_3$ and $F^\bullet$ (○). Relative electrode potential ($\Delta U$) is labelled with arrows.

### $BF_4^-$ anodic oxidation

As follows from Table 1, the total ionic charge on $BF_4^-$ noticeably depends on the coverage, varying from −0.7$e$ to −0.4$e$. Hence, due to the strong inter-ionic repulsion, −$E_{ads}$ decreases with increasing $\theta$ (Fig. 3). Oppositely, the charge on $F^\bullet$ is only slightly above −0.4$e$ at all coverages.[§] Thus, in the absence of strong inter-ionic repulsion, $E_{ads}(F^\bullet)$ weakly depends on $\theta$ (Fig. 3).

At $\theta \geq 1/3$, when a $BF_4^-$ anion is flipped, it spontaneously dissociates to $BF_3$ and $F^\bullet$ (Fig. 1b). Notably, that according to Gutsev et al.[83] "$BF_4^\bullet$ has to be considered rather as a van der Waals complex". Accordingly, the observed dissociation is a consequence of $BF_4^-$ oxidation at anodic polarisation: 1) $BF_4^-$ = $e^-$ + $BF_4^\bullet$; 2) $BF_4^\bullet$ = $BF_3$ + $F^\bullet$. The formed fluoride $F^\bullet$ can further recombine to $F_2$ and react with RTIL components, as was observed in work [102], or form a covalent bond with surface atoms, for instance, forming $MeF_n$, as described in Ref. [103]. The formed $BF_4^-$ can also form a stable $B_2F_7^-$ complex. Using in situ infrared spectroscopy technique, Romann et al.[61] found that combination of $BF_3$ with $BF_4^-$ anion happens if there are no better electron pair donors. The complex formation shifts the equilibrium further towards the anion breakdown.

**Relative electrode potential and integral capacitance**

DFT calculations capture the oxidation process that sets the anodic limit on the electrochemical window for EMImBF$_4$. Due to the simplicity of the Helmholtz model, the corresponding potential remains overestimated. Table 2 shows the potential values at the Au(111) | $BF_4^-$ interface (MD/DFT) and the polarised Au(111) | EMImBF$_4$ interface (MD). One can see that the MD and DFT results are in agreement for the Au(111) | $BF_4^-$ interface at $\theta < 1/3$. In the MD simulations, the spontaneous flip of anions happens at $\theta = 2/5$ leading to the formation of the second layer of anions and crowding. At such high coverages, in the DFT calculations, $BF_4^-$ loses its charge and decomposes (Table 1). Besides that, both DFT and MD computations show that the distance of closest approach of $BF_4^-$ to the surface decreases with the increase of coverage as well as surface charge, and potential – this is reflected by the slight growth of the integral capacitance shown in Fig. 4.

Fig. 4a shows the integral capacitance values calculated according to Eqs. 4–7 for the systems with all $BF_4^-$ ions in the same orientation, i.e. characterised by the largest possible interfacial dipole moment. The match between the Eqs. 4–7 curves implies consistency in the mechanism and energetics: the accumulation of ions happens via simple physical adsorption, and the integral energy rises due to repulsion between counter-ions.

The integral capacitance for the more complex multilayer Au(111) | EMImBF$_4$ interface shows the impact of the ionic layering on the potential magnitude. In Fig 4b, at the same potential value, the capacitance is higher for the multilayer model than for the Helmholtz model. At the same coverage (or surface charge) the relative electrode potential values are smaller for the multilayer model than for the Helmholtz model, as shown in Table 2. Only when the monolayer of counter-ions of maximal density is formed, the potential, surface charge, and integral capacitance are the same for both models. In our simulations, this takes place at the coverage of 2/5 at 6.3 V. Above this value crowding of anions occurs.

The obtained capacitance values are in reasonable agreement with experimental values of 5–10 μF/cm² for high-frequency differential capacitance measured at gold single crystal (100) and (111) surfaces.[28,29,40,43,64–66] However, the Helmholtz model capacitance increases with increasing the relative electrode potential, while the multilayer layer model capacitance decreases in the same way as in experimental work [104].

**Table 2.** Relative electrode potential ($\Delta U$) values calculated for different surface coverage using MD simulations of Helmholtz (MD$_H$) and multilayer (MD$_{ML}$) models in comparison to the DFT data.

| $\theta$ | DFT | MD$_H$ | MD$_{ML}$ |
|---|---|---|---|
| 1/3 | 4.6 V | 5.3 V | 4.3 V |
| 1/4 | 3.8 V | 4.1 V | 2.7 V |
| 1/6 | 3.3 V | 2.9 V | 1.3 V |
| 1/12 | 2.2 V | 1.7 V | 0.4 V |
| 1/20 | 1.4 V | 1.1 V | 0.15 V |

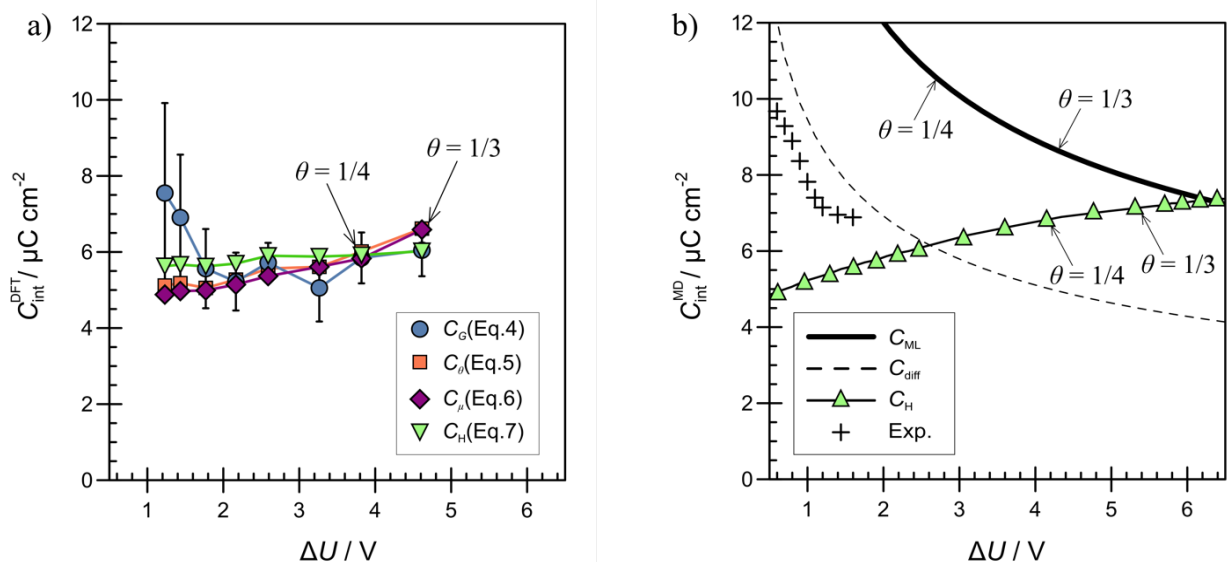

**Figure 4.** a) Integral capacitance dependence on potential, calculated using the DFT data and Eqs. 4–7. b) Integral ($C_{ML}$ and $C_H$) and differential ($C_{diff}$) capacitance dependencies on potential calculated using the MD data. Surface coverage values ($\theta$) are labelled with arrows. Experimental data (Exp.) taken from Ref. [104], where Au | EMImBF$_4$ interface was studied using electrochemical impedance spectroscopy.

**Anionic adlayers vs dense monolayer of anions**

The presented results can be interpreted through theories developed by Loth et al., Bazant–Storee–Kornyshev or Yochelis, as well as molecular-level interpretations by Feng et al. and Ivaništšev et al.[18,20,21,27,105] According to Feng et al., the multilayer structure can be divided into layers with zero net charge (except for the innermost layer) and characterised by alternating dipole moments from layer to layer.[21] The presence of the structured RTIL above the innermost layer decreases the absolute value of the potential drop across the interface. That is why in Fig. 4 the integral capacitance for the multilayer model is higher than for the Helmholtz model. According to Ivaništšev et al., the capacitances are equal at the potential of the monolayer formation when a single monolayer of counter ions completely compensates the surface charge.[27] These potentials manifest the transition from overscreening to crowding regimes.[18,27,106] At lower absolute potentials, due to the anion–cation correlation, there is an alternating layer of anions and cations; at higher absolute potentials, the surface is crowded by the counter-ions.

In the presented MD simulations, the potential of the monolayer formation was found to be 6.3 V (Fig. 4b, the cross point of $C_H$ and $C_{ML}$), which corresponds to the coverage of 4/5 or 65 μC/cm$^2$ (accounting for polarizability of ions). However, the DFT calculations demonstrated that at a comparable potential scale the interface becomes unstable already at 3.5 V ($\theta = 1/3$, Figs. 1b and 3). In a hypothetical DFT based MD simulations of the multilayer model, the decomposition potential should be even lower, firstly, due to the dissociation of the oxidised $BF_4^-$, and, secondly, due to the ionic layering. In experiment, anodic electrochemical reactions start around 1.6 V vs PZC.[104]

Notice, that from a geometrical point of view the $BF_4^-$ patterns at the coverage higher than 1/6 completely occupy the innermost layer, as the free space on the surface is sterically hindered for larger cations. In the MD simulations of the multilayer model, this coverage corresponds to 1.3 V (Table 2). Thus, we surmise that similar structures were visualized by the scanning tunnelling microscopy in works [66,67], where the formation of ordered anionic adlayers was assigned to low voltages of less than 1 V vs PZC. We stress that in this case, the observation of the anionic adlayers does not rule out the overscreening nor the presence of EMIm$^+$ cations next to the innermost layer. Moreover, the accumulation of anions starting from the coverage of 1/6 and ending with the monolayer formation at 4/5 requires marked potential and surface charge increase. Consequently, 1) anionic adlayers and the dense monolayer of anions are different structures, 2) the adlayers are part of the multilayer EDL, and 3) the crowding of anions at low potential values is extremely improbable.

**Helmholtz vs Multilayer models**

Comparison of the DFT and MD results provides a ground for resolving whether the EDL in RTILs is *one*- or *multi*-ionic-layer thick. As expected, for the *one*-ionic-layer thick, according to Helmholtz model (Eq. 7), the capacitance is almost independent of the potential. For the multilayer model, both integral and differential capacitance decreases with increasing the relative electrode potential (Fig. 4b). The same tendency was shown for the differential capacitance of the Au | EMImBF$_4$ interface in the experiment.[104] Also, previous computations by Feng et al.[21] and Hu et al.[107] showed that while some qualitative trends might be captured by structural changes in the innermost layer, the subsequent layers have an essential influence in defining the dependence of capacitance on electrode potential. All in all, the multilayer EDL in EMImBF$_4$, as a whole, determines the overall potential-dependent capacitance.

In the Helmholtz model, the EDL thickness is defined as the distance of closest approach on counter-ions to the surface. Only for this model, the positions of the ionic charge and the ionic mass planes coincide. The first one defines the potential drop in the corresponding parallel plate capacitor. The second appears due to steric effects. Both might be equal to the counter-ion radius under an assumption that the surface charge plane lies at a surface-atom-radius distance the surface plane (defined by the nuclei positions).

For the multilayer structure, the ionic charge and mass plane positions are different. The ionic mass density is positive at any distance from the surface, while the sign of the ionic charge density ($\rho$) depends on the excess of anions or cations at a given distance from the surface ($z$). The ionic charge plane position ($z_{ion}$) is expressed as:

$$z_{ion} = -\int \rho(z) z \, dz / \sigma. \quad (9)$$

$z_{ion}$ value can be smaller than the counter-ion radii. On the contrary, in the multilayer structure, any *i*-th layer lies further from the surface than of the innermost layer.

Force–distance curves, as those provided in works [49,108], can be used to count the number of layers and to estimate the geometrical thickness of the EDL. However, such thickness would be useless for calculating the EDL capacitance as the surface charge–potential dependence is dictated by the ionic charge plane position. To clarify the difference let us simplify the multilayer into an ionic bilayer.

**Ionic bilayer model**

In the recent ionic-bilayer model, the multilayer was presented as a bilayer – the contact layer of counter-ions and the subsequent layer of co-ions.[109] It is a simplified representation of the multilayer EDL. The integral and differential capacitances of this model are given as:

$$C = \varepsilon\varepsilon_0/(d - \delta\lambda/\sigma), \quad (10)$$

$$C_{diff} = \varepsilon\varepsilon_0/(d - \delta(d\lambda/d\sigma)), \quad (11)$$

where $\delta$ is the geometrical distance between the first and the second layers, $\varepsilon$ is the high-frequency dielectric constant, and $\lambda$ is the co-ion charge density in the second layer that is equal in magnitude, but opposite in sign to the surface charge excess of the counter-ions in the first layer.[106]

Let us use this simple model to reflect the thickness of the EDL. Note, condition $\lambda = 0$ means that the model simplifies to the Helmholtz model. Otherwise, by definition, the EDL in the ionic-bilayer model is two-layer with a constant width of $(d + \delta)$. Herewith, for the integral capacitance, the denominator in Eq. 10 is smaller than $d$. That is why the integral capacitance is higher in the simulations of the multilayer model than of the Helmholtz model (Fig. 4b). For the differential capacitance, the denominator in Eq. 11 might be equal to $(d + \delta)$ only when the surface charging relies solely on the exclusion of the co-ions from the second layer to the bulk, i.e. $d\lambda = -d\sigma$. Competition with the more traditional mechanism of charging, i.e. adsorption of counter-ions on the surface, ensures that the condition $d\lambda = -d\sigma$ is not satisfied. Yet, the change in the rate of the second layer destruction ($d\lambda/d\sigma \rightarrow -1$) causes the decrease of the capacitance with increasing the potential after a maximal co-ion charge density is accumulated in the second layer when $d\lambda/d\sigma = 0$.[109] From Fig. 4b one can deduce that this occurs at an intersection of $C_H$ and $C_{diff}$ around 2.6 V. Note, the capacitance decreases with increasing potential also at lower potentials. According to Eq. 11, the capacitance peak (not shown in Fig. 4b) appears at a potential when the accumulation of co-ions is maximal ($d\lambda/d\sigma \rightarrow d/\delta$), and the capacitance inevitably decreases above this potential. Most important for our discussion is that the geometric thickness of the ionic bilayer model is unrelated to its capacitance vs potential dependence.

As soon as there are two or more charged layers in the EDL, one should account at least for two layers in the geometrical interpretation of the potential-dependent capacitance. The knowledge of the PZC position is essential. It allows for using both Eqs. 10 and 11 by converting the differential capacitance to the integral capacitance or surface charge–potential dependence. In principle, one can estimate $\delta$, $\lambda$, and $d\lambda/d\sigma$ values, among which the only $\delta$ is a geometric parameter.

For a qualitative example consider works [34,110,111], in which based on Eq. 7 with $\varepsilon = 8$ authors concluded that $d$ values for the same anion vary in monocationic and dicationic RTILs: $d_{mono} > d_{di}$. The same authors highlighted the simplicity of the Helmholtz-type models as well as the need for a more advanced theory. From Eq. 10 a more expectable relation emerges: $\delta_{mono} < \delta_{di}$, meaning that the position of the dication layer lies further from the surface than of the monocation layer while the position of the anionic layer remains the same.

The given alternative explanation relies on hypotheses that call for new experimental and computational studies. A more detailed analysis of published experimental results is also desirable.[104,112] The generic hypothesis is that the ionic charge plane position could be estimated by accounting for two ionic layers, as in the ionic bilayer model.[109] Substituting the one-ionic-layer thick foundation of the Helmholtz model should be a step forward a more general model of the EDL in RTILs.

## Conclusions

We have studied the adsorption of $BF_4^-$ anions from 1-ethyl-3-methylimidazolium tetrafluoroborate ($EMImBF_4$) on the charged Au(111) surface using DFT and MD computations. The study represents a crucial piece of the scientific puzzle. It addresses the question: does the innermost layer dominate in the overall potential-dependent multilayer EDL? It also illustrates how the relative electrode potential is set through the interfacial potential drop across the model interface, and how it serves as a measuring tape for adsorption and oxidation of $BF_4^-$ anions.

First, DFT calculations and MD simulations of the simplest *Helmholtz* model of the Au(111) | $BF_4^-$ interface give similar results, once a surface change plane is accounted in the MD simulations. The agreement holds up to high anodic potentials, wherein the DFT calculations $BF_4^-$ spontaneously oxidizes: $BF_4^- = e^- + F^\bullet + BF_3$. The Helmholtz capacitance is almost independent of the potential.

Second, in the MD simulations of the multilayer model of the polarised Au(111) | $EMImBF_4$ interface, the potential drop is significantly reduced due to the ionic layers above the innermost layer. For the multilayer model, the capacitance is dependent on the potential, in-line with the experimental results.[104]

We conclude that the multilayer EDL in $EMImBF_4$, as a whole, determines the overall potential-dependent capacitance. Consequently, we suggest that the Helmholtz model (Eq. 7) should be used with great caution in the case of RTILs, as the EDL in ionic liquids is apparently *multi*-ionic-layer thick. We recommend accounting for the multilayer EDL structure when discussing properties of RTIL-based energy storage and transformation devices, especially supercapacitors.

## Conflicts of interest

There are no conflicts to declare.

## Acknowledgements

This work was supported by the EU through the European Regional Development Fund under project TK141 "Advanced materials and high-technology devices for energy recuperation systems" (2020.4.01.15-0011), by the Estonian Research Council (institutional research grant No. IUT20-13), by the Estonian Personal Research Projects PUT1107 and PUT360, and by short-term scientific missions funded by COST actions MP1303 and CM1206. This work was also supported by Graduate School of Functional materials and technologies receiving funding from the European Regional Development Fund at the University of Tartu, Estonia. The financial support from Fundação para a Ciência e Tecnologia (FCT/MEC) funds, and co-financed by the European Union (FEDER funds) under the Partnership Agreement PT2020, through projects POCI/01/0145/FEDER/007265 with reference UID/QUI/50006/2013, and POCI/01/0145/FEDER/006980 with reference UID/QUI/UI0081/2013 (LAQV@REQUIMTE and CIQUP, respectively) and postdoctoral research grant SFRH/BPD/97918/2013 is also acknowledged. Results were obtained in part using the High-Performance Computing Center of the University of Tartu, in part using the EPSRC funded ARCHIE-WeSt High-Performance Computer (www.archie-west.ac.uk, EPSRC grant no. EP/K000586/1).

We thank Enn Lust for inspirational discussions of the new phenomena in RTILs at electrified interfaces. We firmly acknowledge the contribution of Maxim V. Fedorov in the interpretation of the presented results. Last but not least, we are grateful to Renat R. Nazmutdinov for his consultations in DFT calculations.

## Notes and references

† Adsorption – is an increase in the concentration of a substance at the interface of a liquid phase due to the operation of surface forces. In the most cases studied, the *physical* adsorption of $BF_4^-$ arise due to Coulomb attraction of ions to the charged gold surface. The *specific* adsorption $F^{\bullet}$ takes place only when $BF_4^-$ anion oxidizes.

‡ Amazon EC2 real-time computing resources were accessed by using the D2CM tool, designed to migrate a complete software environment from a local desktop directly to the cloud.[80,81] D2CM tool allowed us to scale calculations on-demand without queues, which are common for traditional High-Performance Computation facilities. It also enabled us to define the full lifecycle of the calculations by specifying which input files and scripts to include, which commands to execute and which files to download when a calculation ends. Overall, it has significantly simplified the use of elastic cloud computing resources for performing electronic structure calculations.

§ For comparison, in $AuF_3$ molecule, the average calculated charge on fluorine is −0.35e. Additional analysis suggests that Au–F in $AuF_3$ is a polar covalent. At the modelled Au(111) surface, fluorine adsorbs preferably at the hollow site, *i.e.* coordinating with three gold atoms. For the sake of simplicity, we denote the adsorbed fluorine as a radical $F^{\bullet}$.